# Scaling optical computing in synthetic frequency dimension using integrated cavity acousto-optics

Han Zhao[1*]†, Bingzhao Li[1]†, Huan Li[1], and Mo Li[1,2*]

[1]Department of Electrical and Computer Engineering, University of Washington, Seattle, WA 98195, USA

[2]Department of Physics, University of Washington, Seattle, WA 98195, USA

†These authors contributed equally to the work

*Email: hzhao89@uw.edu; moli96@uw.edu

ABSTRACT

Optical computing with integrated photonics brings a pivotal paradigm shift to data-intensive computing technologies. However, the scaling of on-chip photonic architectures using spatially distributed schemes faces the challenge imposed by the fundamental limit of integration density. Synthetic dimensions of light offer the opportunity to extend the length of operand vectors within a single photonic component. Here, we show that large-scale, complex-valued matrix-vector multiplications on synthetic frequency lattices can be performed using an ultra-efficient, silicon-based nanophotonic cavity acousto-optic modulator. By harnessing the resonantly enhanced strong electro-optomechanical coupling, we achieve, in a single such modulator, the full-range phase-coherent frequency conversions across the entire synthetic lattice, which constitute a fully connected linear computing layer. Our demonstrations open up the route towards the experimental realizations of frequency-domain integrated optical computing systems simultaneously featuring very large-scale data processing and small device footprints.

Analog optical computing encodes and processes data using continuously variable quantities of light [1-3]. While optical nonlinearity requires high power expense, linear optical components can perform data movement, temporal-spatial signal processing and multiply-accumulate operations with potentially unparalleled bandwidth, speed and energy efficiency [4-7]. As the current digital electronic computing technologies approach the physical limit, such advantages of optics motivate the recent development in building optical accelerators that can sustain the ever-growing data demand at the hardware level [8-16]. Integrated photonics provides a powerful optical computing platform that benefits from scalable fabrications and integration compatibility with electronic circuits, affording architectures with rapid programmability [11-16]. Considerable progress has been made in building integrated photonic neural networks with high data throughput by incorporating time and/or wavelength division multiplexing [15,16]. However, realizing large-scale, fully connected networks on photonic chips can be very challenging. Most $N \times N$ optical computing layers based on spatial encoding require $O(N^2)$ scaling of photonic components, occupying huge device footprints compared to the electronic counterparts. Such footprint-inefficient scaling poses a major roadblock for integrated photonic computing from being applied in some important architectures such as the multilayer perceptron (MLP) [17].

The emerging notion of synthetic frequency dimension provides a promising strategy to drastically scale up the optical computing systems in both classical and quantum regimes [18-25]. Encoding information as coherent optical fields on a synthetic frequency lattice increases the fan-in/fan-out of a single photonic logic unit, thus improving the scalability of data processing by orders of magnitudes. The implementations of frequency-domain $N \times N$ optical networks require efficient modulators that simultaneously link the $N$ discrete nodes via coherent frequency conversions [19,25]. For this purpose, integrated acousto-optic modulators can stand out with high modulation efficiency and large modulation depth by exploiting the strong optomechanical interaction between co-localized optical and acoustic modes [26-31]. Recent thin-film lithium niobate modulators have reached modulation depth that can couple a few sidebands [29,31]. Nonetheless, a single device that can compose a fully connected computing layer on a sizable synthetic frequency lattice remains unrealized. Achieving the most efficient on-chip acousto-optic modulation requires simultaneously optimized optomechanical coupling and piezoelectric transduction on a monolithic material platform. To this end, the heterogeneous integration of silicon on insulator (SOI) with complementary metal-oxide-semiconductor (CMOS)-compatible piezoelectric materials such as aluminum nitride (AlN) holds promise for high-performance integrated modulators [32-34], which will offer the key building blocks for data-intensive frequency-domain optical computing systems.

Here, we demonstrate scalable matrix-vector multiplications (MVM) in the synthetic frequency dimension by leveraging an efficient nanophotonic cavity acousto-optic modulator on the AlN-on-SOI platform. The very large dynamic modulation depth arising from the engineered strong electro-optomechanical coupling enables the coherent frequency conversions among a myriad of sidebands spanning a synthetic frequency lattice. Thereby, with a single such modulator, we realize a large-scale, fully connected computing layer that performs linear transformations on the complex-valued vector inputs encoded as spectrally coherent optical fields (Fig. 1a). We highlight the advantage of the persistent long-range spectral phase coherence of the MVM operations performed by our modulator. Our device contributes the critical component to a highly scalable and hardware-efficient integrated photonic computing architecture based on concatenated layers of modulators.

Fig. 1b shows a scanning electron microscopy (SEM) image of our device fabricated on an AlN-on-SOI substrate [35]. The modulator consists of a one-dimensional photonic crystal cavity etched in a suspended silicon rib waveguide which is connected to the AlN/Si piezoelectric region with a silicon sleeve area in between. The nanophotonic cavity is end-coupled to a pair of grating couplers for optical input/output. We achieved a high loaded quality factor $Q_L = 4.6 \times 10^5$, corresponding to a total cavity loss rate of $\kappa = (2\pi) \cdot 420$ MHz (Fig. 1c). Acoustic waves are excited by driving a split-finger interdigital transducer (IDT) patterned on the free-standing AlN/Si region with an RF signal, and subsequently propagate to the optical waveguide via the silicon membrane. The IDT is designed with a period of 3 μm to excite a set of mechanical modes with angular frequencies $\Omega > \kappa$, reaching the sideband-resolved regime (Fig. 2a) [35]. By etching a free-edge reflector on the lower side of the waveguide, we create an acoustic resonator that coherently builds up a strong mechanical displacement field at the nanophotonic cavity. Such mechanical motion effectively modulates the optical resonance through a combination of moving-boundary and photoelastic effects [36]. Under the modulation, the intra-cavity photon dynamics can be described as

$$\dot{a}(t) = [i(\Delta - \beta \cdot \hat{f}(t)) - \kappa/2]a(t) + \sqrt{\kappa_{ex}}\, a_{in}(t), \quad (1)$$

where $\kappa_{ex}$ is the external coupling rate, $a_{in}(t)$ is the input optical field, and $\Delta = \omega_p - \omega_0$ is the detuning of the input laser (angular) frequency $\omega_p$ from the cavity center frequency $\omega_0$. $\hat{f}(t)$ denotes the normalized acoustic waveform. $\beta = 2g_{om}/\Omega$ is the modulation index that measures the dynamic modulation depth, where $g_{om}$ is the optomechanical coupling proportional to the amplitude of the mechanical mode.

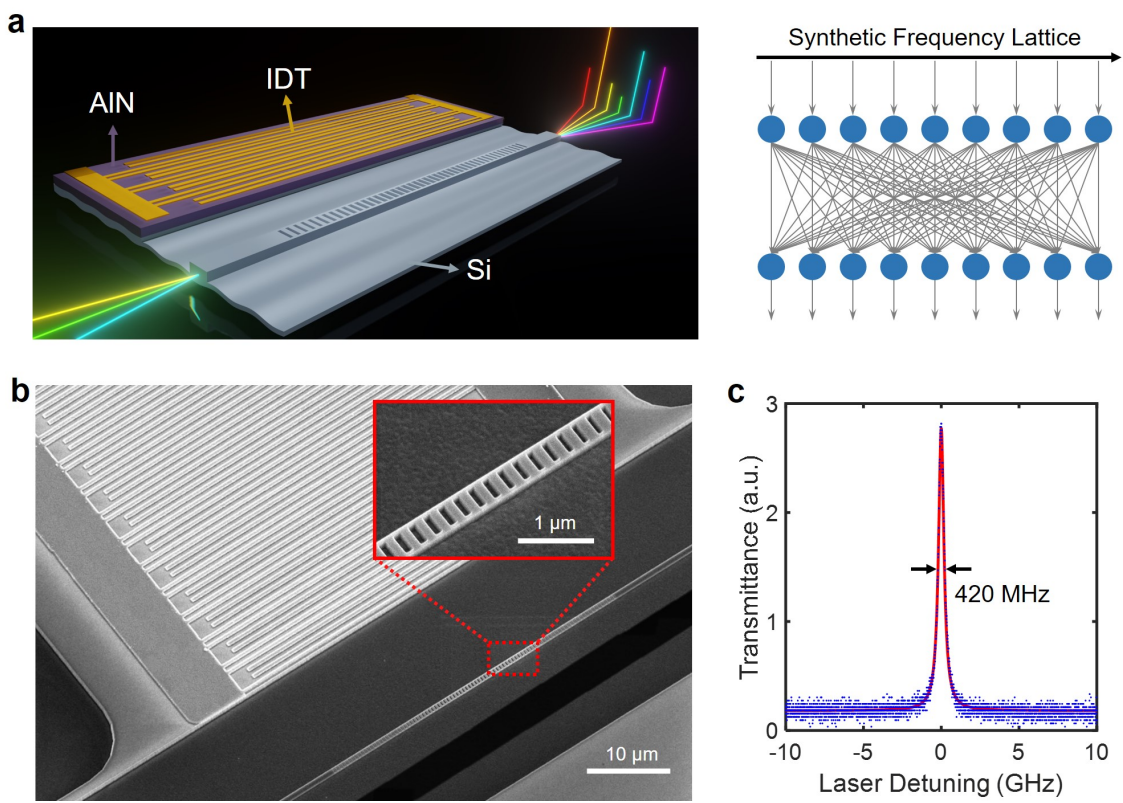

**Fig. 1. Nanophotonic cavity acousto-optic modulator that performs scalable matrix-vector multiplications in the synthetic frequency dimension.** (**a**) Schematic illustration of our acousto-optic modulator, which is equivalent to a fully connected linear optical computing layer on the synthetic frequency lattice. (**b**) SEM image of the modulator fabricated on the AlN-on-SOI platform. Inset is the zoom-in image of the nanophotonic cavity. (**c**) Measured optical transmission spectrum without modulation, showing a linewidth of 420 MHz. The center optical resonance wavelength is around 1547.5 nm.

Our acousto-optic modulator features ultra-high modulation efficiency in the sideband-resolved regime. The resulting deep modulation can generate multiple resolved sidebands at $\omega_0 \pm s\Omega$ with integer sideband order $s$, forming a synthetic lattice in the frequency domain. To quantify the modulation efficiency and the achievable size of the synthetic lattice, we drive the IDT at the mechanical resonances and measure the optical transmission spectra with varying RF power [35]. Fig. 2b shows the results of the 2.903 GHz drive ($\Omega/\kappa \sim 7$), from which we infer a characteristic half-wave voltage $V_\pi = 580$ mV for $\beta = \pi$ by fitting the measurements with the theory [35]. At an RF power of 3 dBm ($\beta = 2.41$), we observe the emergence of multiple sidebands up to $\pm$3rd orders, comprising a finite lattice of 7 sites. The 803 MHz drive ($\Omega/\kappa \sim 2$) can induce much more efficient modulation with the lowest $V_\pi = 19$ mV (Fig. 2c). We obtain the maximal $\beta = 22.9$ at an RF power of -7 dBm, which generates a synthetic lattice of approximately 50 sites over a wide frequency range of 40 GHz.

The centerpiece of performing fully connected MVM with our modulator is the coherent conversions from each input frequency site to all the sites at the output. To understand this, we consider a monochromatic laser input and an RF drive with a single microwave tone $\hat{f}(t) = \cos(\Omega t + \phi)$. At a large $\beta$, the incident photons can absorb or emit multiple phonons because of the strong optomechanical coupling. Consequently, the input optical field is scattered to a set of harmonic signals $\{a_l\}$ detuned from the cavity center frequency by $\Delta + l\Omega$, where $l$ is the harmonic order. By solving Eq. (1), it can be derived that

$$a_l = \sum_k J_{l+k}(\beta) J_k(\beta) e^{-in\phi} \frac{\kappa_{ex} a_{in}}{i(-\Delta + k\Omega) + \kappa/2}, \quad (2)$$

where $J_\nu(x)$ is the $\nu$-th order Bessel function of the first kind [35]. We perform heterodyne measurements to characterize the amplitudes of all the harmonic signals with varying input laser frequency. Fig. 2d and Fig. 2e show the exemplary optical spectra of the harmonics measured at the RF drives of 2.903 GHz and 803 MHz, respectively, highly consistent with the theoretical values [35]. When the input laser frequency is set on the $n$-th synthetic lattice sites, i.e., detuned by $\Delta = n\Omega$, each $a_l$ leads to a non-local frequency conversion from the $n$-th site to the $m$-th site, where $m = n + l$. Hence, the entire set of harmonic generations at all sidebands constitute a two-dimensional optomechanical coupling tensor

$$\mathbf{G} = [g_{mn}], \, m,n \in [-M,M]: \, g_{mn} = a_{m-n}/a_{in}, \quad (3)$$

where $2M + 1$ is the size of the synthetic lattice determined by the modulation index $\beta$. More generally, for an optical input vector on the synthetic lattice $\mathbf{x} = (x_{-M},...,x_0,...,x_M)^T$, our modulator performs a complex-valued MVM $\mathbf{y} = \mathbf{G} \cdot \mathbf{x}$, yielding an output vector $\mathbf{y} = (y_{-M},...y_0,...,y_M)^T$. We highlight that, with $\beta_{max} = 22.9$, a single such modulator provides an MVM unit with a scalable size of up to 50×50 in the frequency domain.

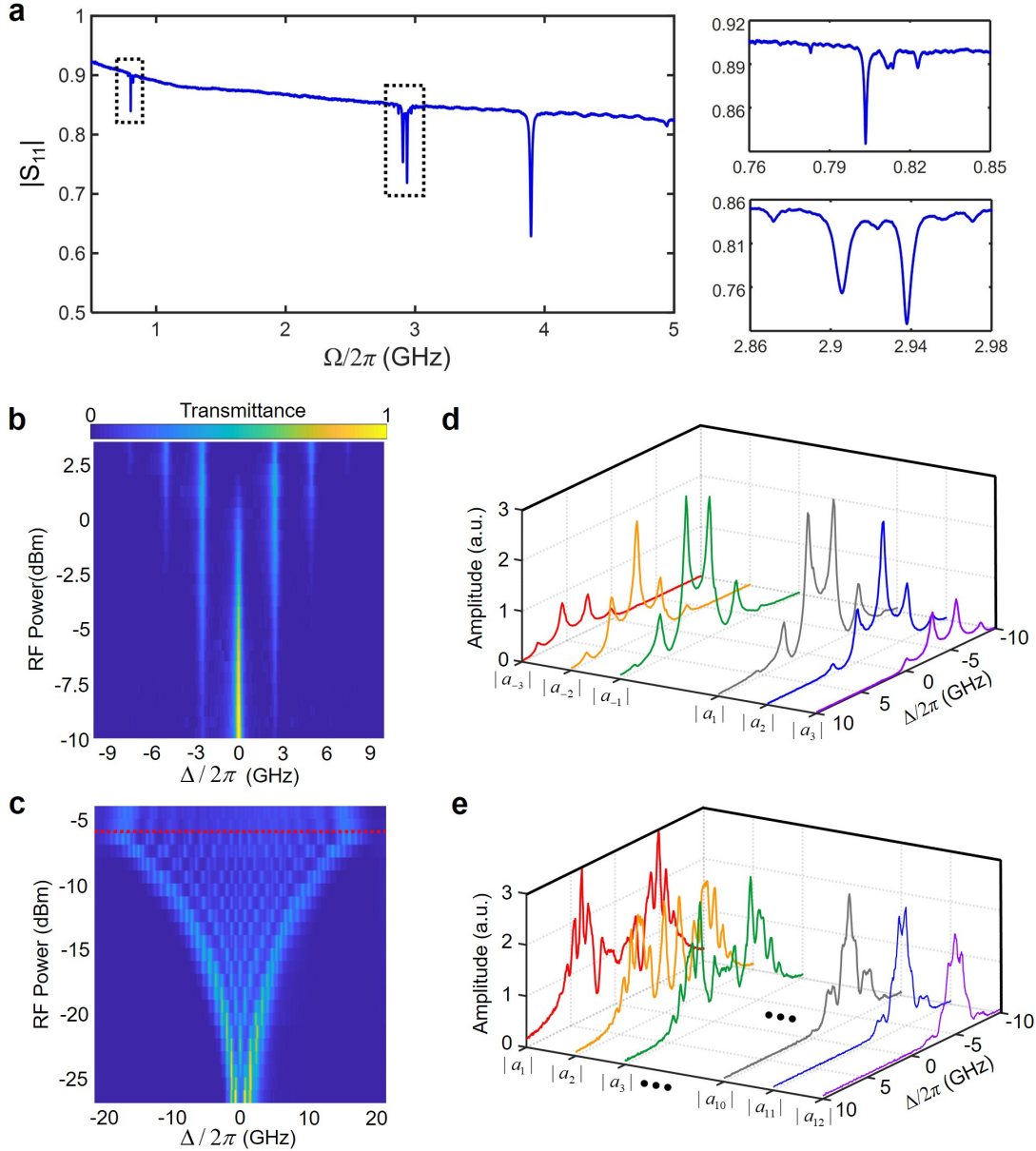

**Fig. 2. Characterizations of the acousto-optic modulation efficiency and the harmonic signal generations.** (**a**) Reflection (amplitude) spectrum of the IDT. The mechanical resonances in the zoom-in spectra at ~800 MHz and ~2.90 GHz have pronounced modulation efficiency, whereas the resonance at 3.95

GHz has a negligibly weak modulation effect. The star and diamond marks denote the prominent frequencies (803 MHz and 2.903 GHz) with enhanced modulation efficiencies for the corresponding mechanical modes [35]. (**b**) Optical transmission spectra under the modulation driven at 2.903 GHz with varying RF power from -10 dBm to 3 dBm. (**c**) Optical transmission spectra under the modulation driven at 803 MHz with RF power from -27 dBm to -5 dBm. The red dashed line denotes the threshold of an electromechanical nonlinearity above which the number of sidebands ceases increasing. (**d**) Measured optical spectra of the non-vanishing harmonic signal amplitudes $|a_l|$ ($l = \pm 1, \pm 2, \pm 3$) at 2.903 GHz, -2.5 dBm RF drive ($\beta = 1.29$). (**e**) Spectra of the positive-order harmonic signal amplitudes $|a_l|$ ($l = 1, 2, 3, …, 10, 11, 12$) at 803 MHz, -17 dBm RF drive ($\beta = 6.90$). The higher-order harmonics decrease significantly beyond $l = 12$ at this drive. The corresponding spectra of the negative-order harmonics (not plotted) are mirror-symmetric about $\Delta = 0$.

In addition to the high scalability, another outstanding advantage of the frequency-domain MVM is the persistent phase coherence across the entire synthetic lattice. In contrast to conventional spatial-domain schemes, which are susceptible to various causes of decoherence such as device defects, non-uniformity and thermal fluctuations, the phase information transmitted through the synthetic lattice is intrinsically preserved by the coherent photon-phonon interactions in our modulator. To demonstrate the scalable and coherent MVM, we operate our device using the $\Omega = 803$ MHz drive and set $\beta = 11.3$, which generates a 25×25 matrix **G** according to Eq. (3). A Mach-Zehnder intensity modulator **M**$_\text{I}$ is used to synthesize a vector input of three coherent frequency components, including the carrier transmission and the two opposite-sign sideband signals with their complex amplitudes controlled by a DC bias and an RF drive at $\Omega$, respectively. For simplicity, we tune the temporal delay of the modulations to be zero ($\phi = 0$) [35]. The output on the synthetic frequency lattice is thereby a result of the weighted complex-number summation of the corresponding columns in **G**, representing the coherent MVM operation.

For the first MVM experiment, we set the laser frequency at the center optical resonance ($\Delta = 0$), and remove the carrier transmission, thus providing an input vector $\mathbf{x} = (...,0, x_{-1}, 0, x_1, 0,...)^\text{T}$ where $x_{-1} = -x_1$ (Fig. 3a). These two input components couple to the 25 sites on the synthetic lattice through the non-local frequency conversions, which add up to a symmetric amplitude distribution at the output (Fig. 3b). Next, we tune the DC bias to generate an input $\mathbf{x} = (...,0, x_{-1}, x_0, x_1, 0,...)^\text{T}$, where $|x_0| = |x_1|$ and $\arg(x_0) = -\arcsin(x_1/a_\text{in})$ (Fig. 3c). Such additional component $x_0$ is also coherently scattered to the entire frequency

lattice and induces the interference with the pattern formed by $x_{\pm 1}$, leading to an asymmetry in Fig. 3d. To demonstrate the coherence of the full range "edge-to-edge" connections, we then align the laser frequency to the 9th site ($\Delta = 9\Omega$), and set $\mathbf{M_I}$ to produce an input $\mathbf{x} = (...,0, x_8, x_9, x_{10}, 0,...)^T$ with $|x_8| = |x_9| = |x_{10}|$. The DC bias on $\mathbf{M_I}$ is switched between the phase relation of either $\arg(x_9) = \arcsin(x_{10}/a_{in})$ (Fig. 3e) or $\arg(x_9) = -\arcsin(x_{10}/a_{in})$ (Fig. 3g). From the measured output amplitude patterns, we observe the suppression (Fig. 3f) and revival (Fig. 3h) of the negative-order sites near the opposite edge by only phase-flipping of the carrier transmission, a strong indicator of the built-in long-range phase coherence with the frequency-domain MVM operations.

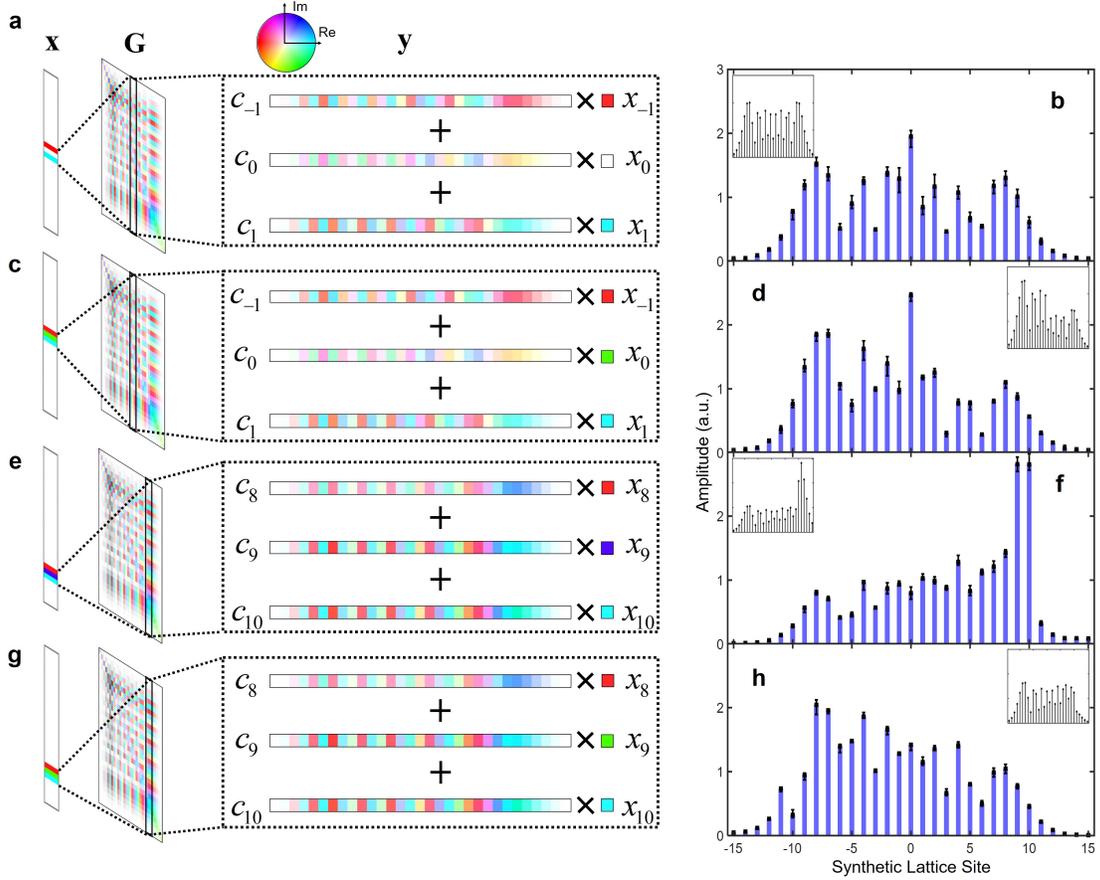

**Fig. 3. Large-scale coherent matrix-vector multiplications in the synthetic frequency dimension.** The large matrix $\mathbf{G}$ is configured by driving our acousto-optic modulator at 803 MHz with -13 dBm power ($\beta = 11.3$). The vector input $\mathbf{x}$ of three coherent frequency components is synthesized and controlled by an intensity modulator $\mathbf{M_I}$. (**a**) MVM with laser frequency at the 0th synthetic lattice site ($\Delta = 0$) and $\mathbf{M_I}$ set at suppressed carrier transmission. (**c**) $\Delta = 0$ and $\mathbf{M_I}$ set with three equal-amplitude components and a negative phase on the carrier transmission. (**e**) and (**f**) $\Delta = 9\Omega$ and $\mathbf{M_I}$ set with three equal-amplitude components and positive/negative phases on the carrier transmission. (**b**), (**d**), (**f**), (**h**) Measured output amplitudes on the synthetic lattice from the settings in (a), (c), (e) and (g), respectively. $c_j$ represents the *j*-th column of **G**, and $x_k$ denotes the input at the *k*-th lattice site. Error bars are calculated from 5 measurements. Insets in (b), (d), (f) and (h) are the corresponding results from the theoretical calculations. The discrepancies between the measurements and the theory primarily come from the background transmission of the optical cavity, which is not included in the theoretical model.

Practical optical computations using integrated photonics require programmability with a high degree of freedom to allow on-chip optimization processes such as backpropagation training in a neural network. For the frequency-domain computing architecture, a straightforward approach to increase independently tunable parameters is to concatenate multiple modulators controlled by separate electronics. To this end, we notice the general phase modulations in the full parametric space, including the broadband elements, constitute a non-abelian (noncommutative) group $<G,\cdot>$, where the cascading of modulators defines the binary operation " $\cdot$ " with the matrix-matrix multiplication. The noncommutativity can manifest as nonreciprocal frequency conversions resulted from the coupling phase anisotropy and the non-unitarity [35]. To demonstrate the feasibility of the concatenation architecture, we implement two elements of the group $\mathbf{G}, \mathbf{M} \in <G,\cdot>$ with our device and a broadband electro-optic modulator (EOM), driven by the same microwave tone. We cascade them in both $\mathbf{G} \cdot \mathbf{M}$ (Fig. 4a) and $\mathbf{M} \cdot \mathbf{G}$ (Fig. 4b) orders and probe the responses with a laser frequency $\Delta = 0$. Since broadband EOMs typically have a low modulation efficiency with $V_\pi$ at a few volts, we choose the RF drive at 2.903 GHz and add a 20-dB amplifier to the driving arm of the EOM to reach comparable modulation indices. In this scheme, the RF driving phases at both **G** and **M** count as independently programmable parameters in addition to the modulation depth. To reflect this increased programmability, we use a tunable RF phase shifter which controls the temporal delay between the two modulations. Fig. 4c compares the measured output amplitudes for both the concatenation orders with varying modulation phase delay. The contrast of the results clearly shows the driving phase control of the noncommutative frequency conversions, featuring the non-abelian algebraic structure.

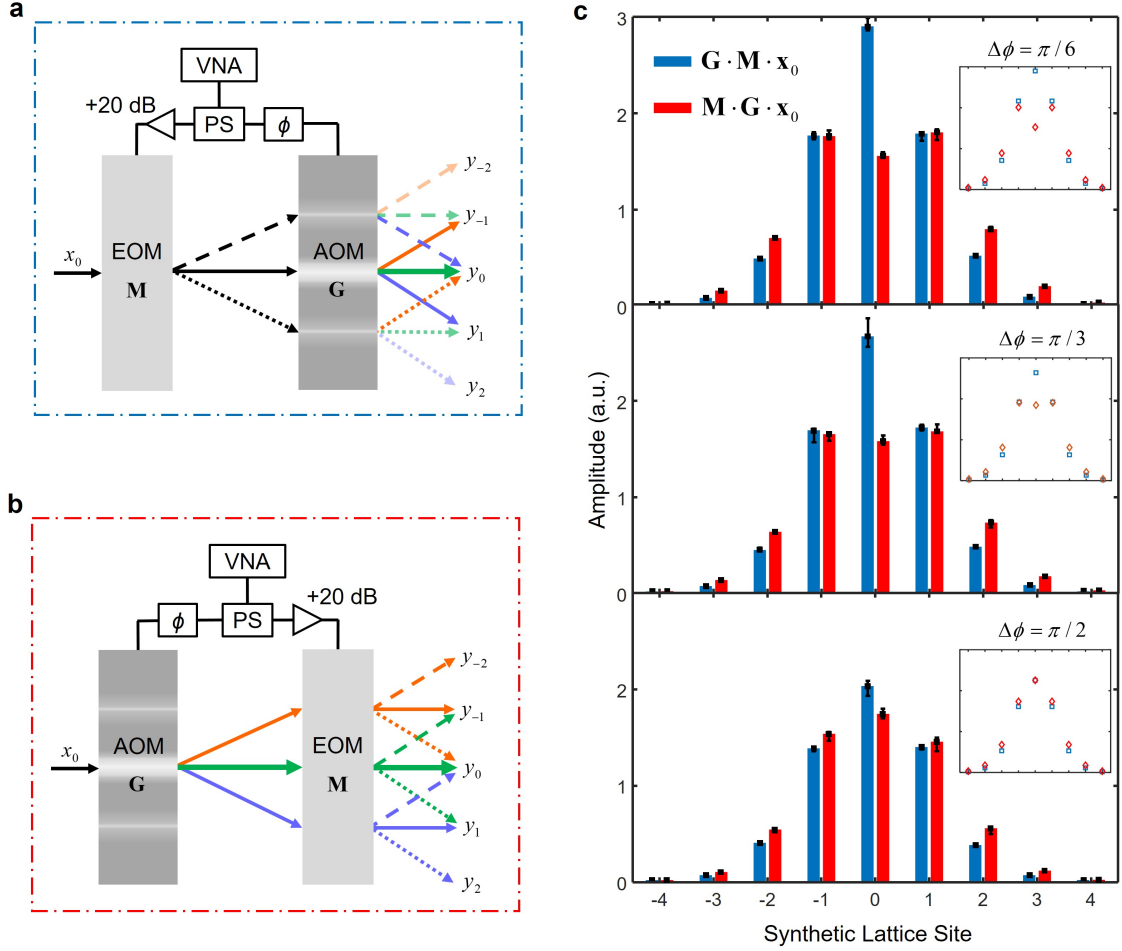

**Fig. 4. Concatenations of phase modulators.** (**a**) Setup of concatenated broadband electro-optic phase modulator (EOM, **M**) and our acousto-optic modulator (AOM, **G**) driven by the same vector network analyzer (VNA). A laser input at $\Delta = 0$ ($\mathbf{x}_0$) goes through the EOM first and then the AOM, corresponding to $\mathbf{G}\cdot\mathbf{M}\cdot\mathbf{x}_0$. (**b**) Setup of the concatenation in the reverse order, $\mathbf{M}\cdot\mathbf{G}\cdot\mathbf{x}_0$. Orange, green and blue lines represent the -1st, 0th (carrier) and 1st harmonic signals generated by the AOM, respectively. Higher-order harmonics are neglected for the illustration. Solid, dotted and dashed lines represent the carrier and ±1st sideband signals by the EOM. $y_k$ denotes the output of the $k$-th synthetic lattice sites. PS: RF power splitter; +20 dB: RF amplifier with 20 dB gain; $\phi$: RF phase shifter. (**c**) Contrast of the output between the two concatenation orders with varying phase delay $\Delta\phi$ at the two modulations. Error bars are plotted from 5 measurements. Insets are the corresponding results from the theoretical calculations.

In conclusion, we have demonstrated an ultra-efficient nanophotonic acousto-optic modulator on the hybrid AlN-on-SOI platform that performs large-scale, complex-valued MVM in a fully connected fashion. We reveal the advantages of high scalability and long-range phase coherence associated with the MVM in the synthetic frequency dimension.

Further considerations on the engineering of the electromechanical transducer will allow the generation of multiple harmonic tones at $n\Omega$ [37,38], which facilitates more hardware-efficient realizations towards arbitrary MVM operations. Our demonstrations open the door to a disruptively new silicon-based optical computing architecture that is scalable with significantly increased operand vector length in compact footprints. Such frequency-domain scheme also extends naturally to scalable quantum computing by incorporating optical quantum sources such as single-photon parametric down-conversion, spontaneous four-wave mixing [39-40], and microwave qubit transduction [33], where the efficient dynamic modulators facilitate frequency-bin logic gates for high-dimensional photonic qudits [41].

**References**

1. Prucnal, P. R. & Shastri, B. J. *Neuromorphic photonics*. (CRC Press, 2017).
2. Caulfield, H. J. & Dolev, S. Why future supercomputing requires optics. *Nat. Photonics* **4**, 261-263 (2010).
3. Solli, D. R. & Jalali, B. Analog optical computing. *Nat. Photonics* **9**, 704-706 (2015).
4. Liu, W. et al. A fully reconfigurable photonic integrated signal processor. *Nat. Photonics* **10**, 190-195 (2016).
5. Nahmias, M. A. et al. Photonic multiply-accumulate operations for neural networks. *IEEE J. Sel. Top. Quantum Electron.* **26**, 7701518 (2019).
6. Hamerly, R. et al. Large-scale optical neural networks based on photoelectric multiplication. *Phys. Rev. X.* **9**, 021032 (2019).
7. Estakhri, N. M., Edwards, B. & Engheta, N. Inverse-designed metastructures that solve equations. *Science* **363**, 1333-1338 (2019).
8. Lin, X. et al. All-optical machine learning using diffractive deep neural networks. *Science* **361**, 1004-1008 (2018).
9. Spall, J., Guo, X., Barrett, T. D. & Lvovsky A. I. Fully reconfigurable coherent optical vector–matrix multiplication. *Opt. Lett.* **45**, 5752-5755 (2020).
10. Zhou, T. et al. Large-scale neuromorphic optoelectronic computing with a reconfigurable diffractive processing unit. *Nat. Photonics* **15**, 367-373 (2021).

11. Shen, Y. et al. Deep learning with coherent nanophotonic circuits. *Nat. Photonics* **11**, 441-446 (2017).

12. Tait, A. N. et al. Silicon photonic modulator neuron. *Phys. Rev. Applied* **11**, 064043 (2019).

13. Bogaerts, W. et al. Programmable photonic circuits. *Nature* **586**, 207-216 (2020).

14. Zhang, H. et al. An optical neural chip for implementing complex-valued neural network. *Nat. Commun.* **12**, 457 (2021).

15. Xu, X. et al. 11 TOPS photonic convolutional accelerator for optical neural networks. *Nature* **589**, 44-51 (2021).

16. Feldmann, J. et al. Parallel convolutional processing using an integrated photonic tensor core. *Nature* **589**, 52-58 (2021).

17. Liu, H., Dai, Z., So, D. R. & Le, Q. V. Pay Attention to MLPs. Preprint at https://arxiv.org/abs/2105.08050 (2021).

18. Ozawa, T. et al. Synthetic dimensions in integrated photonics: From optical isolation to four-dimensional quantum Hall physics. *Phys. Rev. A* **93**, 043827 (2016).

19. Lukens, J. M. & Lougovski, P. Frequency-encoded photonic qubits for scalable quantum information processing. *Optica* **4**, 8-16 (2017).

20. Yuan, L., Lin, Q., Xiao, M. & Fan, S. Synthetic dimension in photonics. *Optica* **5**, 1396-1405 (2018).

21. Bell, B. A. et al. Spectral photonic lattices with complex long-range coupling. *Optica* **4**, 1433-14336 (2017).

22. Titchener, J. G. et al. Synthetic photonic lattice for single-shot reconstruction of frequency combs. *APL Photonics* **5**, 030805 (2020).

23. Dutt A., et al. A single photonic cavity with two independent physical synthetic dimensions. *Science* **367**, 59-64 (2020).

24. Tusnin, A. K., Tikan, A. M. & Kippenberg, T. J. Nonlinear states and dynamics in a synthetic frequency dimension. *Phys. Rev. A* **102**, 023518 (2020).

25. Buddhiraju, S. et al. Arbitrary linear transformations for photons in the frequency synthetic dimension. *Nat. Commun.* **12**, 2401 (2021).

26. Bochmann, J., Vainsencher, A., Awschalom, D. D. & Cleland, A. N. Nanomechanical coupling between microwave and optical photons. *Nat. Phys.* **9**, 712-716 (2013).

27. Li, H., Tadesse, S. A., Liu, Q. & Li, M. Nanophotonic cavity optomechanics with propagating acoustic waves at frequencies up to 12 GHz. *Optica* **2**, 826-831 (2015).

28. Balram, K. C. et al. Acousto-optic modulation and optoacoustic gating in piezo-optomechanical circuits. *Phys. Rev. Applied* **7**, 024008 (2017).

29. Shao, L. et al. Microwave-to-optical conversion using lithium niobate thin-film acoustic resonators. *Optica* **6**, 1498-1505 (2019).

30. Forsch, M. et al. Microwave-to-optics conversion using a mechanical oscillator in its quantum ground state. *Nat. Phys.* **16**, 69 (2020).

31. Jiang, W. et al., Efficient bidirectional piezo-optomechanical transduction between microwave and optical frequency. *Nat. Commun.* **11**, 1166 (2020).

32. Tian, H. et al., Hybrid integrated photonics using bulk acoustic resonators. *Nat. Commun.* **11**, 3073 (2020).

33. Mirhosseini, M., Sipahigil, A., Kalaee, M. & Painter, O. Superconducting qubit to optical photon transduction. *Nature* **588**, 599-603 (2020).

34. Kittlaus, E. A. et al. Electrically driven acousto-optics and broadband non-reciprocity in silicon photonics. *Nat. Photonics* **15**, 43-52 (2021).

35. See supplementary materials.

36. Rakich, P. T., Davids, P. & Wang, Z. Tailoring optical forces in waveguides through radiation pressure and electrostrictive forces. *Opt. Express* **18**, 14439 (2010).

37. Schülein, F. J. et al. Fourier synthesis of radiofrequency nanomechanical pulses with different shapes. *Nat. Nanotech.* **10**, 512-516 (2015).

38. Weiß, M. et al. Multiharmonic frequency-chirped transducers for surface-acoustic-wave optomechanics. *Phys. Rev. Applied* **9**, 014004 (2018).

39. Kues, M. et al. On-chip generation of high-dimensional entangled quantum states and their coherent control. *Nature* **546**, 622-626 (2017).

40. Wang. J, et al. Multidimensional quantum entanglement with large-scale integrated optics. *Science* **360**, 285-291 (2018).

41. Kues, M. et al. Quantum optical microcombs. *Nat. Photonics* **13,** 170-179 (2019).

**Acknowledgements**

This work was supported by NSF Award EFMA-1741656 and EFMA-1641109. Part of this work was conducted at the Washington Nanofabrication Facility/Molecular Analysis Facility, a National Nanotechnology Coordinated Infrastructure (NNCI) site at the University of Washington with partial support from the National Science Foundation via awards NNCI-1542101 and NNCI-2025489.

**Author contributions**

H. Z. and M. L. conceived the project. H. Z. and B. L. fabricated the sample and performed the experiments. All authors contributed to data analysis and manuscript preparation.

**Supplementary Materials**

1. Fabrication of the cavity acousto-optic modulator on the AlN-on-SOI platform

      Our cavity acousto-optic modulator was fabricated by the processes shown in Fig. S1. The substrate was prepared by sputtering 320-nm thick polycrystalline aluminum nitride (AlN) on silicon-on-insulator (SOI) wafer with 220-nm Si layer (grown by OEM group). Before patterning the structure, a layer of silicon dioxide ($SiO_2$) was deposited as a hard mask by plasma-enhanced chemical vapor deposition (PECVD). We first patterned the window for Si photonic structures by electron-beam lithography (EBL) with positive ZEP520 resist (developer: amyl acetate), which was subsequently transferred to the $SiO_2$ hard mask by fluorine-based inductively-coupled-plasma etching (ICP-F). The exposed AlN in the window was removed by another step of chlorine-based inductively-coupled-plasma etching (ICP-C) with $Cl_2$/$BCl_3$/Ar chemistry followed by nitrogen plasma cleaning. The etching time was precisely controlled to remove only the targeted AlN layer. The Si photonic structures, including the one-dimensional photonic crystal cavity, grating couplers and waveguide, were patterned using aligned EBL and negative hydrogen silsesquioxane (HSQ) resist (developer: tetramethylammonium hydroxide, TMAH), followed by an ICP-C etching with $Cl_2$ plasma. Another round of aligned EBL (resist: ZEP520) and ICP-C was applied to etch through the free-edge reflector and the releasing windows for the final releasing step. The remaining positive resist and on-top oxide (including the remaining HSQ) were removed by N-Methyl-2-pyrrolidone (NMP) and buffered oxide etchant (BOE), respectively. We then patterned the interdigital transducer (IDT) using a third-time aligned EBL (resist: ZEP520), followed by electron-beam evaporation (Evap) of aluminum and lift-off in NMP. Finally, the device was released by vapor hydrofluoric (HF) acid which etched away the buried oxide layer to achieve the suspended structures. The parameters of our fabricated device are listed in Table S1.

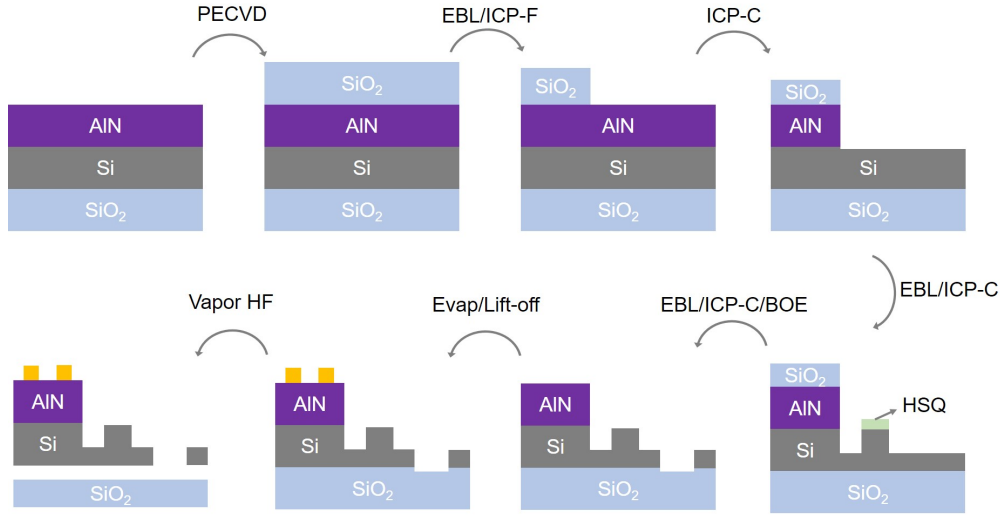

**Fig. S1: Fabrication flow for the cavity acousto-optic modulator on hybrid AlN-on-SOI platform**. PECVD: plasma-enhanced chemical vapor deposition; EBL: electron-beam lithography; ICP-F: fluorine-based inductively-coupled-plasma etching; ICP-C: chlorine-based inductively-coupled-plasma etching; HSQ: hydrogen silsesquioxane; BOE: buffered oxide etchant; Evap: electron-beam evaporation; Vapor HF: vapor hydrofluoric acid.

**Table S1 Parameters of the fabricated cavity acousto-optic modulator**

| | | | |
|---|---|---|---|
| Thickness of AlN (nm) | 320 | Thickness of Si (nm) | 220 |
| Thickness of buried oxide (μm) | 3 | Width of waveguide (nm) | 550 |
| Strip height of waveguide (nm) | 150 | Height of Si sleeve (nm) | 70 |
| Periodicity of photonic crystal (nm) | 350 | Depth of holes (nm) | 150 |
| Length of rectangular holes (nm) | 375 | Maximum hole width (nm) | 180 |
| Minimum hole width (nm) | 80 | Number of tapering | 25 |
| Thicknesses of Al electrodes (nm) | 220 | Periodicity of IDT (μm) | 3 |
| Width of IDT finger (nm) | 375 | Total IDT length (μm) | 150 |
| Distance from IDT to waveguide (μm) | 10 | Distance from waveguide to free-edge reflector (μm) | 2.4 |

2. Analysis of the piezoelectrically transduced mechanical modes

The IDT patterned on the heterogeneous AlN/Si region is used to resonantly excite multiple mechanical modes, which have very distinct acousto-optic modulation efficiencies on the nanophotonic cavity. To understand the relation between the acousto-optic modulations and these mechanical modes, we performed the numerical simulations (COMSOL Multiphysics 5.5) and show in Fig. S2 the displacement fields in the suspended AlN/Si layer, corresponding to the resonances measured by the IDT $S_{11}$ response. The mechanical modes that are of the interest in the main text are the fundamental Lamb mode (large out-of-plane displacement) at ~800 MHz and the fundamental compressional mode (large in-plane displacement) at ~2.9 GHz. Because of the long wavelengths, these two modes strongly couple to the 70-nm Si membrane and the optical nanobeam cavity therefore can induce strong phase modulations. We characterize the acousto-optic modulations by measuring the microwave-to-optical transduction signal $S_{OE}$ in Section 5 and exploit these modes for the frequency-domain matrix-vector multiplications.

Our IDT with the split-finger design can also excite higher-order mechanical modes with odd-number modal orders. In the 500 MHz to 8 GHz spectrum, Mode III is the 3rd-order Lamb mode; Mode V is the 5th-order compressional mode; Mode IV is the 3rd-oder Love mode; and Mode VII is the 3rd-oder AlN/Si Lamb-compression hybrid mode. These higher-order modes are associated with significantly reduced wavelengths which increase the mechanical power dissipation and decrease the modal overlap between the mechanical modes and the optical cavity field (so weaker optomechanical coupling). Consequently, the higher-order mechanical modes have negligible acousto-optic phase modulation efficiencies (compared to the fundamental orders) and dominantly contribute to thermo-optic tuning of the optical cavity resonance. In addition, we also observe the excitations of a symmetric breathing mode (VI). Although this breathing mode shows a larger piezoelectric transduction efficiency, it does not couple to the acoustic wave in the silicon membrane (in Lamb mode) due to the mismatch of the modal symmetry, therefore has no phase modulation effect.

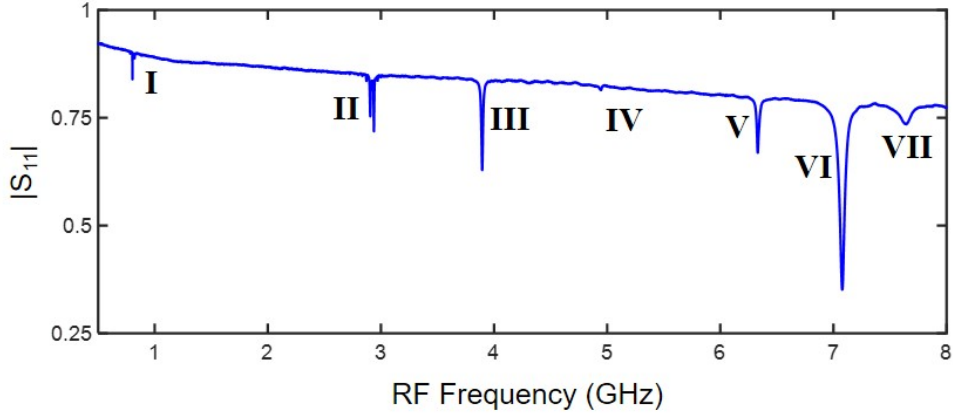

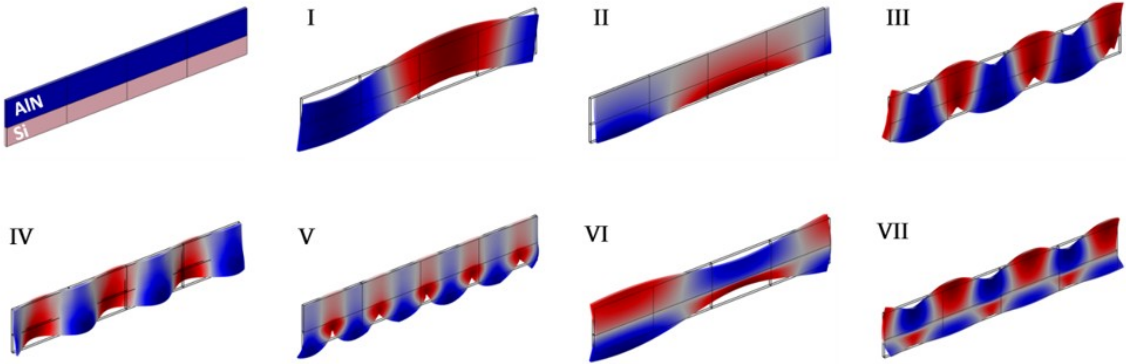

**Fig. S2: Spectrum of the mechanical modes in the AlN/Si region when the IDT is driven in the RF frequency range of 500 MHz – 8 GHz.** I-VII show the simulated distributions of the displacement fields in a single IDT period (3 μm), which correspond to the resonance modes in the experimentally measured IDT spectrum, respectively.

3. Analytical solutions to the intra-cavity photon dynamics

The intra-cavity photon dynamics under the acousto-optic modulation (Eq. (1) in the main text) has explicit solutions when a single microwave tone is applied on the IDT, i.e. $\hat{f}(t) = \cos(\Omega t + \phi)$. The solution provides the insight of the physics under our interrogation. Therefore, before introducing our experimental characterizations, we show here the derivation of the analytical solutions of both the intra-cavity and the output optical fields.

We rewrite the dynamics with $\hat{f}(t) = \cos(\Omega t + \phi)$ as

$$\dot{a}(t) = i\Delta a(t) - i\beta \cdot \cos(\Omega t + \phi)a(t) - \kappa a(t)/2 + \sqrt{\kappa_{\text{ex}}}\, a_{\text{in}}(t), \qquad (1)$$

where $a(t)$ and $a_{\text{in}}(t)$ are the intra-cavity and input optical fields, respectively; $\kappa$ and $\kappa_{\text{ex}}$ denote the total optical cavity decay rate and external coupling rate, respectively; $\Delta = \omega_{\text{p}} - \omega_0$ is the detuning of the input laser angular frequency $\omega_{\text{p}}$ from the cavity center frequency $\omega_0$; $\Omega, \phi$ are the frequency and phase of microwave drive applied to the IDT, respectively; $\beta = 2g_{om}/\Omega$ is the modulation index; and $g_{om}$ is the total optomechanical coupling. By the transformation $a(t) = \alpha(t)\exp[-i\beta\sin(\Omega t + \phi)]$, we obtain

$$\dot{\alpha} = (i\Delta - \kappa/2)\alpha + \sqrt{\kappa_{ex}}e^{i\beta\sin(\Omega t + \phi)}a_{in}. \tag{2}$$

Using Jacobi–Anger expansion

$$\exp[i\beta\sin(\Omega t + \phi)] = \sum_k J_k(\beta)\exp[ik(\Omega t + \phi)], \tag{3}$$

where $J_\nu(x)$ is the Bessel function of the first kind, we decompose Eq. 2 in Fourier series $\alpha(t) = \sum_k \alpha_k \exp(ik\Omega t)$, and obtain

$$ik\Omega\alpha_k(t) = (i\Delta - \kappa/2)\alpha_k + J_k(\beta)e^{ik\phi}a_{in}, \tag{4}$$

which leads to

$$\alpha_k = e^{ik\phi}J_k(\beta)\frac{a_{in}\sqrt{\kappa_{ex}}}{i(-\Delta + k\Omega) + \kappa/2}. \tag{5}$$

The dynamics of the intra-cavity optical field can then be expressed as

$$a(t) = e^{-i\beta\sin(\Omega t + \phi)} \cdot \sum_k \alpha_k \exp(ik\Omega t)$$

$$= \sum_n e^{-in\Omega t}\sum_k J_{n+k}(\beta)J_k(\beta)e^{-in\phi}\frac{\sqrt{\kappa_{ex}}a_{in}}{i(-\Delta + k\Omega) + \kappa/2}. \tag{6}$$

Thereby, the optical output at the exit facet of the end-coupled optical nanobeam cavity can be explicitly calculated as

$$a_{out}(t) = \sum_n e^{-i(\omega_p + n\Omega)t}\sum_k J_{n+k}(\beta)J_k(\beta)e^{-in\phi}\frac{\kappa_{ex}a_{in}}{i(-\Delta + k\Omega) + \kappa/2}. \tag{7}$$

At high modulation index $\beta \gg 0$, the Bessel functions $J_\nu(\beta)$ are non-vanishing for $\nu \gg 1$, leading to large amplitudes at higher order harmonic signals. Hence, the output optical signal for a single-frequency input at $\omega_p$ can be thought of as a compositional baseband of many RF harmonics modulated by the optical carrier frequency $\omega_p$. In the following, we show our method to experimentally characterize each of the RF harmonics in the optical output.

4. Experimental characterization of the cavity acousto-optic modulator

We used homodyne and heterodyne measurement schemes to characterize the modulation in terms of the microwave-to-optical transduction signal ($S_{OE}$) and the harmonic signal generations (Figs. 2-4 in the main text), respectively. Fig. S3 shows the experimental setup. The interdigital transducer (IDT) was driven by the transmitter port (Port 1) of a calibrated vector network analyzer (VNA) with tunable RF frequency and power output. The optical input was realized by coupling a continuous-wave (CW) laser to the on-chip grating coupler through a polarization-maintained fiber, and the output was collected from the output grating coupler by another aligned fiber. We measure the spectra of the direct-current (DC) transmitted optical power (Fig. 1c, Fig. 2b and Fig. 2c) by switching the optical output from our device to a low-speed photodetector (LPD) while sweeping the laser frequency. The high-frequency components of the transmitted optical signal are interrogated by a square-rule high-speed photodetector (HPD) with a bandwidth of 12 GHz, which down-converts the beating notes of the detected optical signal to corresponding RF voltages.

In the homodyne branch, we switched off the acousto-optic frequency shifter (AOFS) and sent the down-converted signal from the HPD to the receiver port (Port 2) of the VNA. The $S_{21}$ parameter of VNA then measured the HPD-generated RF signal at the driving microwave frequency normalized by the input RF complex amplitude, which is proportional to the first-order optical beating note in $a_{out} \cdot a_{out}^*$, where $a_{out}$ is the output field in Eq. (7), i.e.,

$$S_{21} \propto \sum_n a_n \cdot a_{n-1}^*. \tag{8}$$

We note it is only possible for this $S_{21}$ to take nonzero value if the electromechanically transduced acoustic wave modulates the optical field. It is therefore also named the microwave-to-optical transduction signal ($S_{OE} = S_{21}$). In most of previous works where the modulation index is small (only $a_0$, $a_1$ and $a_{-1}$ are relevant), $S_{OE}$ can be simplified to $S_{OE} \propto a_0 a_{-1}^* + a_1 a_0^*$, which is widely used as the metric to the modulation depth and bandwidth when the laser frequency is tuned at the red sideband ($\Delta = -\Omega$, $S_{OE} \propto a_1$) or the blue sideband ($\Delta = \Omega$, $S_{OE} \propto a_{-1}^*$) for sideband-resolved acousto-optic systems. Another functionality of our homodyne measurements is to identify the center optical resonance frequency $\omega_0$ because $S_{OE}$ equals zero at exactly zero detuning $\Delta = 0$ and has a large gradient in the vicinity. We used the traces of the $|S_{21}|$ center local minimum shown in Fig. S4 to characterize the thermo-optic shift induced by the acoustic wave. For our acousto-optic modulator with high modulation index, however, $S_{OE}$ is a complicated composition contributed from the frequency conversions between the adjacent sidebands, and therefore cannot fully characterize the dynamic phase modulation. This necessitates the heterodyne measurements that can spectrally resolve all the harmonic signals received by the HPD.

In the heterodyne branch, we drive the AOFS at an angular frequency $\omega_\mu = (2\pi) \cdot$ 103 MHz, which shifts the optical frequency of the local oscillator (LO) to $\omega_\mu + \omega_p$. When combined with the optical output from our acousto-optic modulator, the signal received at the HPD can be written as (by ignoring the high frequency components)

$$U_{hetero} \propto \left( c_0 e^{-i(\omega_p + \omega_\mu)t} + \sum_n a_n e^{-i(\omega_p + n\Omega)t} \right) \left( c_0 e^{-i(\omega_p + \omega_\mu)t} + \sum_n a_n e^{-i(\omega_p + n\Omega)t} \right)^*. \tag{9}$$

The down-converted RF voltage contains the frequency components at $\omega_\mu - n\Omega$ which have the amplitudes proportional to the corresponding $n$-th harmonic signals in the optical output by a factor of the LO amplitude $h_0$, i.e.

$$|u_{\omega_\mu - n\Omega}| \propto h_0 |a_n|. \tag{10}$$

Therefore, by mapping out all the RF frequency components in a real-time spectrum analyzer (RSA), we can capture all the amplitudes of the harmonic generations induced by the acousto-optic modulation. The heterodyne measurements were used to obtain the experimental results in Figs. 2d, e, Fig. 3 and Fig. 4.

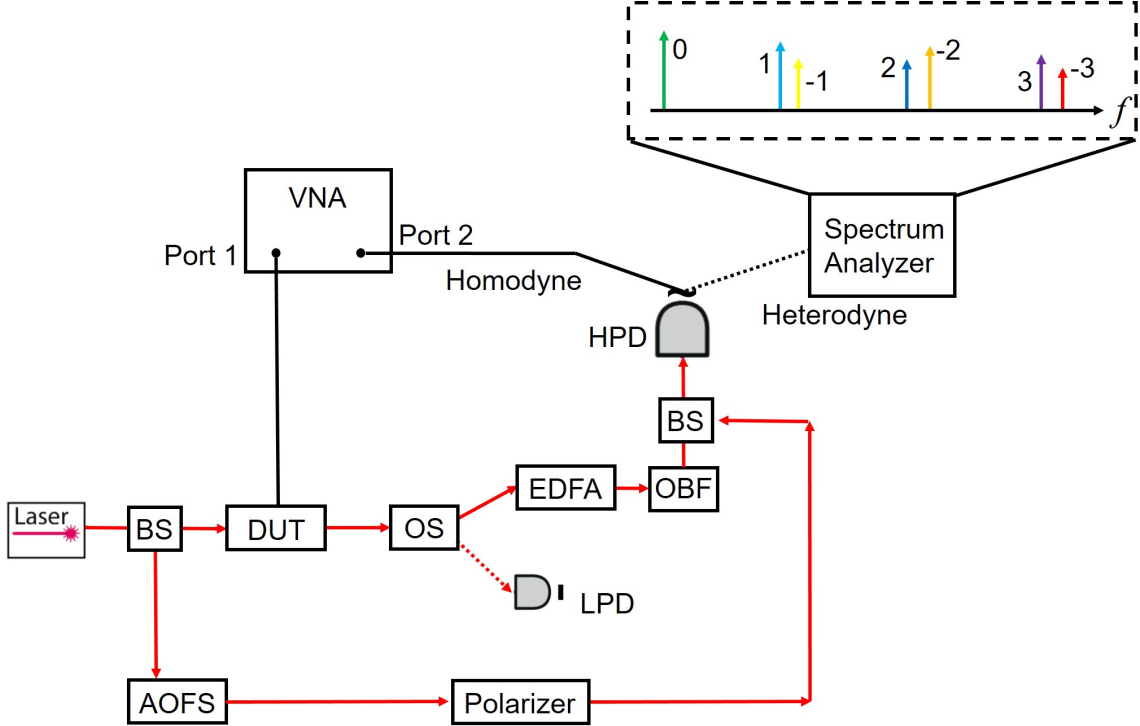

**Fig. S3: Experimental setup for the homodyne/heterodyne measurements**. BS: optical in-line beam splitter; DUT: device under test; AOFS: fiber acousto-optic frequency shifter; OS: optical switch; LPD: low-speed photodetector; EDFA: erbium-doped fiber amplifier; OBF: tunable optical bandpass filter; HPD: high-speed photodetector (12-GHz bandwidth); VNA: vector network analyzer.

5. RF spectra of the microwave-to-optical transduction

We use the homodyne measurements to characterize the spectrum of the acousto-optic modulation and the associated thermo-optic shift for the RF drive at varying RF tone. Fig. S4 shows the spectrum of the measured microwave-to-optical transduction signal at the mechanical resonances, including the prominent fundamental Lamb mode and compressional mode. The acoustic resonator formed by the free-edge reflector gives rise to a series of resonances in the IDT bandwidth. By mapping out the RF spectrum of $S_{OE}$, we were able to identify the on-resonance microwave tones that can induce the most efficient modulation for each mechanical mode. The laser frequency is swept around the intrinsic nanophotonic cavity resonance to probe $S_{OE}$ at all sidebands.

For the fundamental Lamb mode excitation at ~800 MHz, we observed the appearance of $S_{OE}$ at multiple resolved sidebands even at a low RF power of -16dBm. In particular, the 803 MHz drive with an RF bandwidth of 1.3 MHz induces a significantly increased number of sidebands, consistent with the most pronounced electromechanical conversion efficiency measured from $S_{11}$. This RF tone thus facilitates the resonantly enhanced acousto-optic modulation, by which we demonstrated the scalable MVM at a large-scale synthetic frequency lattice. The minimum at zero laser frequency detuning indicates a constant optical center resonance frequency (no pronounced thermo-optic shift) at the -16 dBm RF power (Fig. S4A). For the fundamental compressional mode excitation at ~2.9 GHz, the highest modulation efficiency is achieved at 2.903 GHz, where the optical mode overlaps with the anti-node of the acoustic resonator. The 2.935 GHz resonance has a weaker modulation because the optical mode primarily overlaps with the node of the mechanical standing wave. The acoustic resonances in this frequency range are subject to more mechanical power dissipation, evidenced by the increased linewidth of 7.5 MHz. As a result, a substantial red shift of the optical center resonance frequency can be observed at the RF power of -6 dBm, shown by the trace of the local minimum in the middle of Fig. S4B. As mentioned in Section 2, other higher-order mechanical modes have much reduced modulation efficiency. As an example, we show the spectrum of $S_{OE}$ for the 5th-order compressional mode in Fig. S4D, which is barely measurable even at the RF power of 0 dBm. We remark that the excitations of higher-order mechanical modes, while contributing negligibly to the dynamic phase modulation, can function as thermo-optic resonance tuning, beneficial for aligning the operation frequencies in concatenated modulator networks.

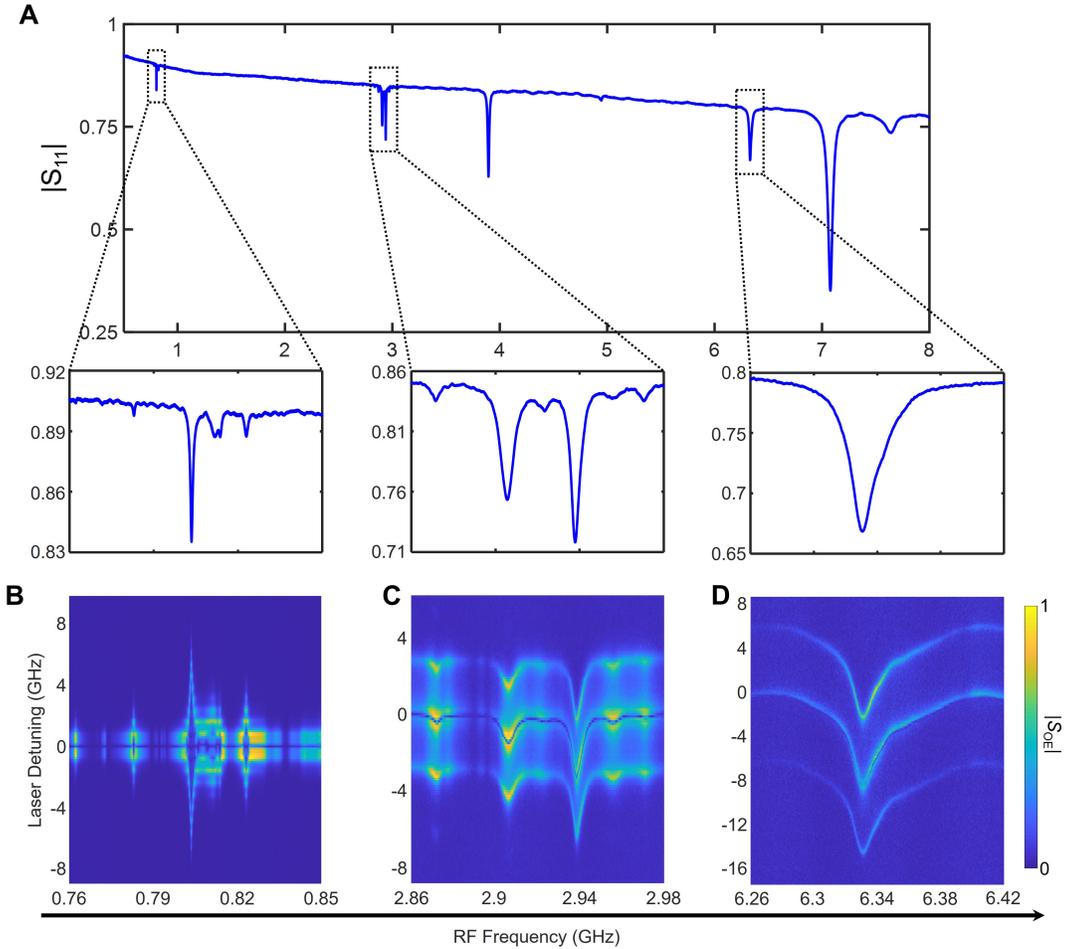

**Fig. S4: Characterizations of the microwave-to-optical transduction**. (**A**) RF spectrum of the IDT reflection coefficient $|S_{11}|$. The zoom-in reflection spectra correspond to the excitations of the fundamental Lamb mode, fundamental compressional mode and 5th-order compressional mode, respectively. (**B**) Amplitude of $S_{OE}$ as a function of the RF driving frequency and the laser detuning for the fundamental Lamb mode. The RF power is fixed at -16 dBm. (**C**) Amplitude of $S_{OE}$ for the fundamental compressional mode at the RF power of -6 dBm. (**D**) Amplitude of $S_{OE}$ for the 5th-order compressional mode at the RF power of 0 dBm.

## 6. Determining the modulation index from the spectra of optical transmittance

One of the consequences of the high acousto-optic modulation index is the generation of multiple sidebands in the optical transmission spectrum. In our sideband resolved system, we can extract the modulation index by fitting the measured spectral features of the split sidebands. To see this, here we show the exemplary fitting results under single microwave tone drives at 2.903 GHz and 803 MHz.

The theoretical values of the DC transmittance can be derived from Eq. (7) and takes the form

$$\langle a_{out} \cdot a_{out}^* \rangle = \sum_n \left| J_n(\beta) \frac{\kappa_{ex} a_{in}}{i(-\Delta + n\Omega) + \kappa/2} \right|^2. \tag{11}$$

We use Eq. (11) to fit the measured spectra of the optical transmittance at varying RF driving power (Figs. 2B,C in the main text), where the parameters $\kappa$, $\kappa_{ex}$, $\Omega$, $a_{in}$ were fixed and only the modulation index $\beta$ is varied to reproduce the spectral features. Fig. S5 shows the fit of a transmittance spectrum at $\Omega$ = 2.903 GHz and 2 dBm RF power. From the agreement between the measured and calculated spectra, we deduced a modulation index $\beta$ = 2.15.

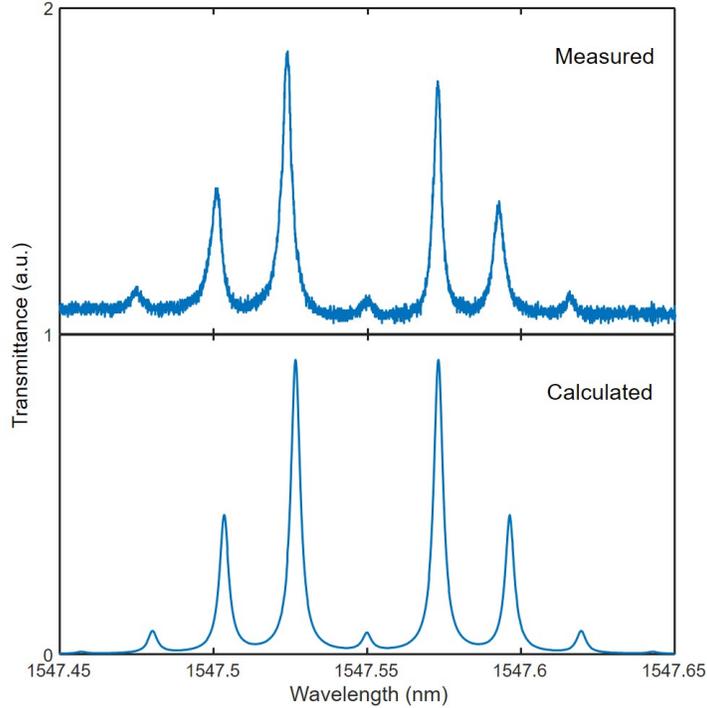

**Fig. S5: Measured optical transmittance spectrum and the corresponding fit curve with $\beta$ = 2.15.** The microwave drive is at $\Omega$ = 2.903 GHz and 2 dBm RF power (upper). The fit curve is calculated by Eq. (11) (lower). The measured spectrum is shifted up by 1 unit.

For large modulation depth observed at $\Omega$ = 803 MHz, the transmission eigenstate distributes to all of the sidebands spanning a wide spectral range and is superposed by the non-uniform background transmission. Nonetheless, we expect a good characterization of the spectral features by a proper fit parameter $\beta$. Fig. S6 shows the fit of the measured transmittance spectrum at $\Omega$ = 803 MHz and -7 dBm RF power. This corresponds to the maximum modulation index of $\beta = 22.9$ obtained before the onset of electromechanical nonlinearity, where the modulation index ceases to increase proportional to the square-root of the RF power.

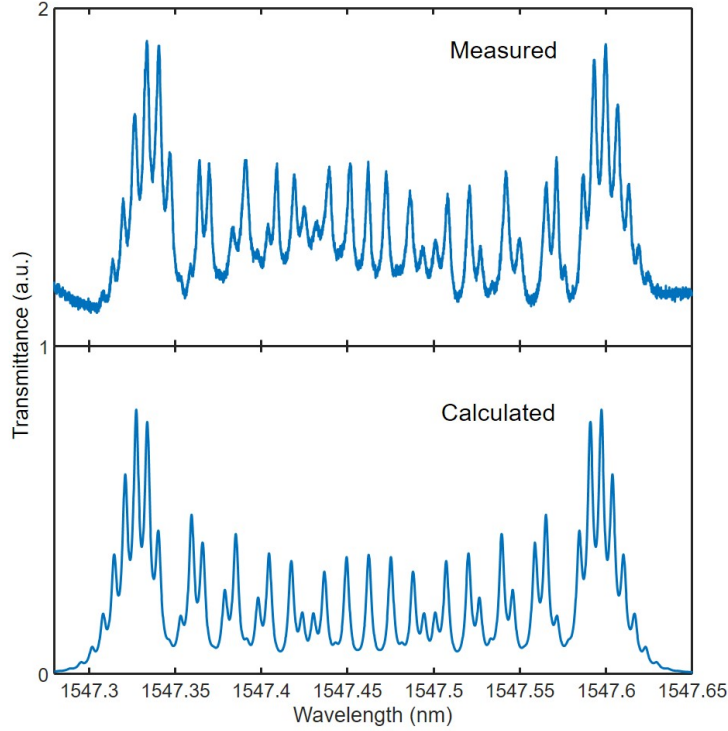

**Fig. S6: Measured optical transmittance spectrum and the corresponding fit curve with $\beta = 22.9$.** The microwave drive is at $\Omega = 803$ MHz and -7 dBm RF power (upper). The fit curve is calculated by Eq. (11) (lower). The measured spectrum is up shifted by 1 unit.

7. Optical spectra of the high-order harmonic signal generations

We used the heterodyne measurement setup to characterize all the harmonic signal generations whose amplitudes are proportional to the corresponding frequency components in the converted RF voltage at the HPD. The theoretical results of the $n$-th harmonic amplitude is

$$|a_n| = \sum_k J_{n+k}(\beta) J_k(\beta) \left| \frac{\kappa_{ex} a_{in}}{i(-\Delta + k\Omega) + \kappa/2} \right|. \tag{12}$$

As we show in Eq. (10), the $n$-th harmonic amplitudes can be experimentally characterized by measuring the heterodyne beating note at the frequency $\omega_\mu - n\Omega$ (with a factor determined by the LO intensity). To reveal the accuracy of our heterodyne characterizations, we show in Fig. S7 the agreement of the measured spectrum of the first-order beating note with the theoretical result by Eq. (12), where $\Omega = 2.903$ MHz and the RF power is -2.5 dBm ($\beta = 1.29$).

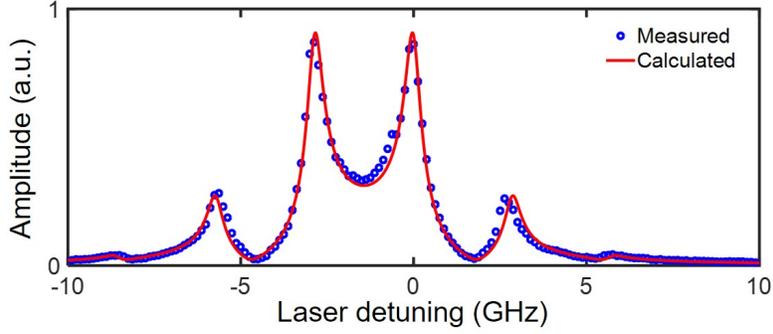

**Fig. S7: Normalized optical spectrum of the first-order beating note obtained at Ω = 2.903 GHz and -2.5 dBm RF power (β = 1.29).**

The characterization scheme also works for large modulation indices observed at Ω = 803 MHz. Within our HPD bandwidth (12 GHz), we show in Fig. S8 the examples of the fit to demonstrate that all the substantial harmonic signals can be read out with high fidelity (β = 6.90).

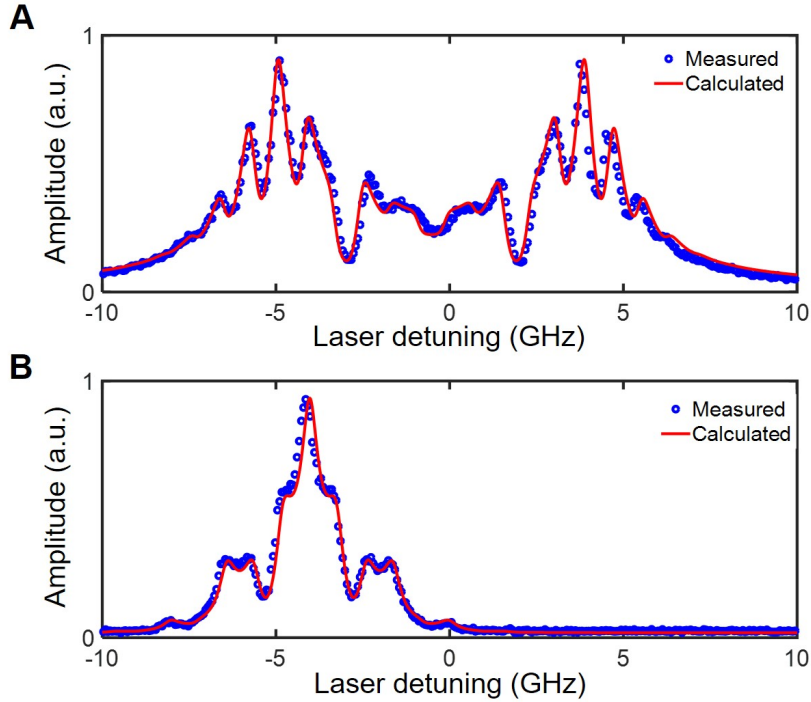

**Fig. S8: Normalized optical spectra of the beating note obtained at Ω = 803 MHz and -17 dBm RF power (β = 6.90). (A)** First-order beating note. **(B)** 10th-order beating note.

8. Driving phase dependence of the matrix-vector multiplications

The optomechanical coupling matrix represented by Eq. (3) in the main text has a dependence on the phase of the RF drive applied on the IDT. Fig. S9 displaces the theoretically calculated dependence on the modulation phase $\phi$ when the device is driven at 803 MHz and with β = 11.3. While this phase variation maintains the amplitudes of the site-to-site couplings ($|g_{mn}|$), a strong phase anisotropy in $g_{mn}$ can be observed, which leads to very different MVM outputs for spectrally coherent vector input. Specifically, for $\phi = 0$, the adjacent columns of **G** have minimum phase contrast, while the long-range coupling phases are considerable. Therefore, in our phase-coherent MVM demonstrations (Fig. 3 in the main text), we chose to set $\phi = 0$ so as to emphasize the persistence of the phase information of the long-range couplings.

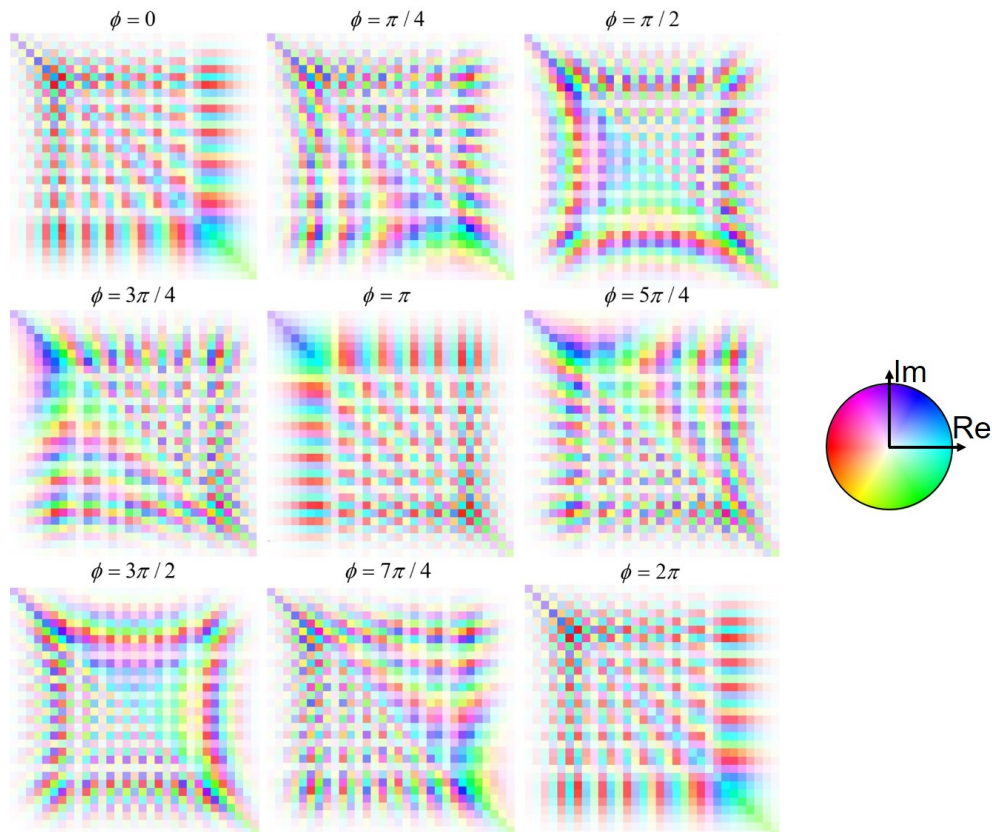

**Fig. S9: Optomechanical coupling matrices at various microwave driving phase.** The complex-valued matrix entries are represented by the false-color mapping. Matrix dimension is 30×30 at $\Omega$ = 803 MHz and $\beta = 11.3$.

9. Measuring the output of the MVM operations on the synthetic frequency lattice

The output amplitude at each frequency site, resulted from the MVM operations (Fig. 3 in the main text), was interrogated by the heterodyne measurements as we have described in Section 4. The experimental setup for the large-scale MVM operations is shown in Fig. S10. The LO frequency $\omega_p + \omega_\mu$ is controlled by the tunable CW laser. When $\omega_p$ is set at one of the frequency sites ($\Delta = s\Omega$), the amplitude of the harmonic signal $|a_l|$ captured by the spectrum analyzer then corresponds to the amplitude at the synthetic lattice site of the order $s + l$. Therefore, we read out the output amplitudes at the synthetic frequency lattice by measuring all non-vanishing harmonic signals. We note, with a photodetector bandwidth of 12 GHz, we were able to fully interrogate up to 15th-order harmonic signals with high fidelity. The harmonic signals at even higher order are subject to decreased detection efficiencies, which set an upper bound for the size of the synthetic lattice in our experimental demonstrations. By using a photodetector with higher bandwidth (> 40 GHz), our system can experimentally realize coherent 50×50 MVM operations with high-fidelity readout.

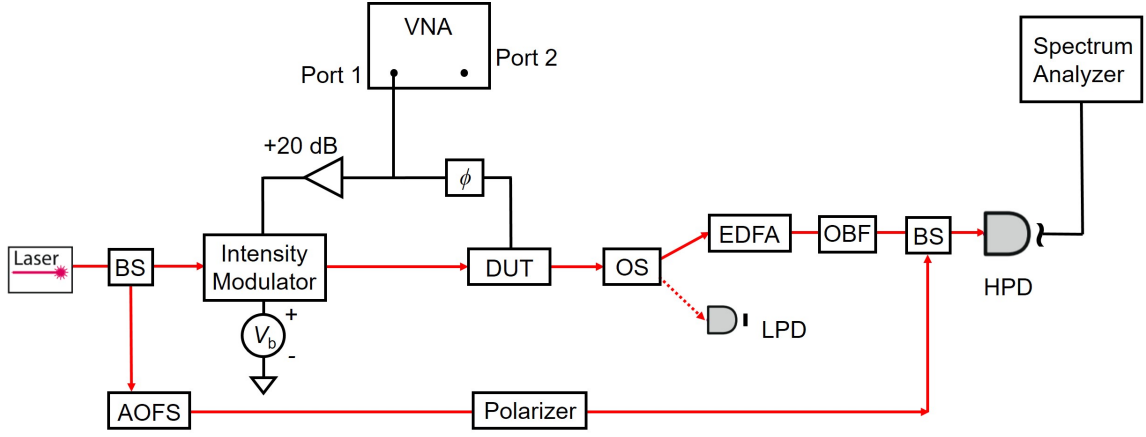

**Fig. S10: Experimental setup of the large-scale MVM.** The intensity modulator, which is driven by the VNA and a DC bias $V_b$, is cascaded before our device to provide a vector input of three spectrally coherent components. The RF phase shifter, $\phi$, is used to tune the modulation phase difference between the intensity modulator and our device, which can be monitored by the transmittance spectrum measurements by the LPD. An RF amplifier of 20-dB gain is applied to drive the intensity modulator in order to obtain pronounced sideband input frequency components for our device.

10. Interrogating the noncommutativity of concatenated modulators

Fig. S11 shows the experimental setups in which we realized the concatenations of our nanophotonic cavity acousto-optic modulator (**G**) with a fiber-coupled electro-optic phase modulator (**M**), in both **G·M** and **M·G** orders. We assume that the modulation phases of the RF drives **G** and **M** are $\phi_1$ and $\phi_2$, and the optical delay between the two modulators is $\tau$. Under the same RF driving tone $\Omega$, the modulation waveforms for the **M·G** order are

$$\begin{cases} \hat{f}_\mathbf{M}(t) = \cos(\Omega t + \phi_2) \\ \hat{f}_\mathbf{G}(t) = \cos[\Omega(t+\tau) + \phi_1] \end{cases}, \quad (13)$$

whereas for the reverse order (**G·M**), the modulation waveforms are

$$\begin{cases} \hat{f}_\mathbf{M}'(t) = \cos[\Omega(t+\tau) + \phi_2'] \\ \hat{f}_\mathbf{G}'(t) = \cos(\Omega t + \phi_1') \end{cases}. \quad (14)$$

We define the modulation phase differences $\Delta\phi = \arg\{\hat{f}_\mathbf{G}(t)\} - \arg\{\hat{f}_\mathbf{M}(t)\} = \Omega\tau + \phi_1 - \phi_2$ and $\Delta\phi' = \arg\{\hat{f}_\mathbf{G}'(t)\} - \arg\{\hat{f}_\mathbf{M}'(t)\} = \phi_1' - \phi_2' - \Omega\tau$. In our experiments, we controlled $\phi_1 - \phi_2$ ($\phi_1' - \phi_2'$) by the RF phase shifter with a tunable phase range $[0, 2\pi]$. We calibrated the optical phase delay $\tau$ to unify the two modulation phase differences associated with the two concatenation orders, by comparing the phase dependences of output amplitudes with the theoretical values.

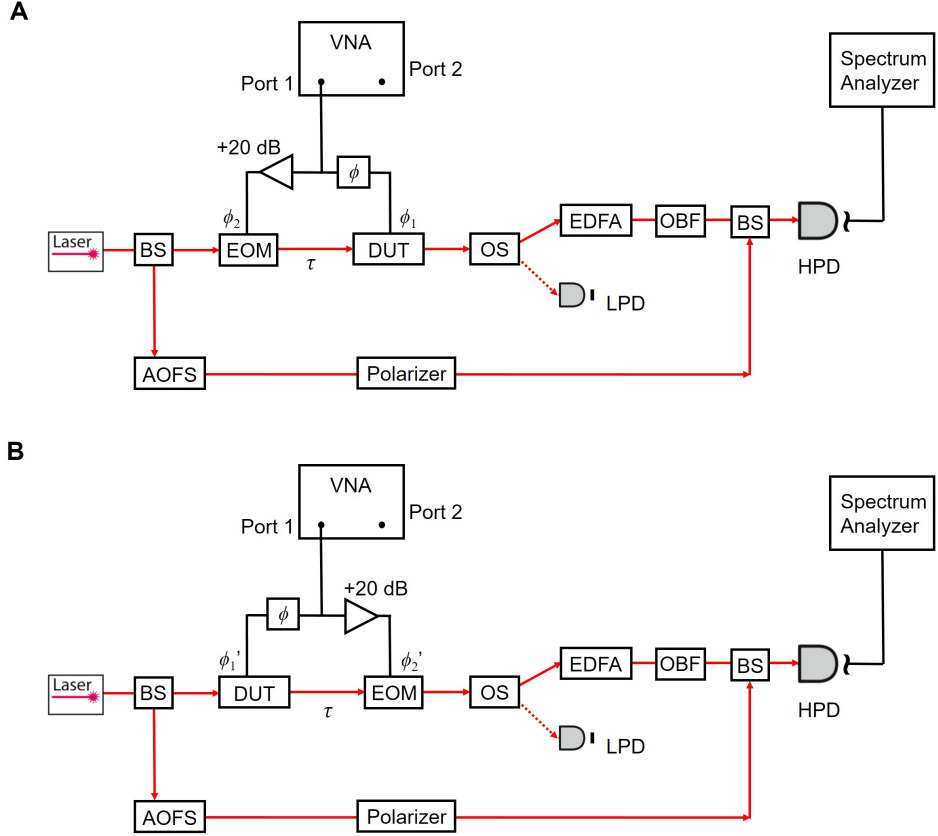

**Fig. S11: Experimental setup of the large-scale MVM.** The intensity modulator, which is driven by the VNA and a DC bias $V_b$, is cascaded before our device to provide a vector input of three spectrally coherent components. The RF phase shifter, $\phi$, is used to tune the modulation phase difference between the intensity modulator and our device, which can be monitored by the transmittance spectrum measurements by the LPD.

To understand the noncommutativity of the non-abelian group formed by cascaded phase modulators, we analyze the matrix-matrix multiplications between **M**: $[m_{kl}]$ and **G**: $[g_{kl}]$. The matrix of a broadband modulator is tri-diagonal represented by the carrier transmission on the diagonal entries $m_{kk}$ and the two opposite-sign sideband generations $m_{k(k-1)}$ and $m_{k(k+1)}$ on the off-diagonal entries ($\phi_2 = 0$). The matrix of our acousto-optic modulator is represented by Eq. (3) in the main text. For simplicity of the analytical calculations, we assume a moderate modulation index that only produces 1st order harmonics ($|k - l| < 2$). With a laser input at $\Delta = 0$, the resulted output vector through $\mathbf{y} = \mathbf{G} \cdot \mathbf{M} \cdot \mathbf{x}_0$ can be expressed as

$$\begin{aligned} y_{-2} &= m_{-1,0} g_{-2,-1} \\ y_{-1} &= m_{0,0} g_{-1,0} + m_{-1,0} g_{-1,-1} \\ y_0 &= m_{-1,0} g_{0,-1} + m_{0,0} g_{0,0} + m_{1,0} g_{0,1} \\ y_1 &= m_{0,0} g_{1,0} + m_{1,0} g_{1,1} \\ y_2 &= m_{1,0} g_{2,1} \end{aligned} \qquad (15)$$

For comparison, the output vector through $\mathbf{y'} = \mathbf{M} \cdot \mathbf{G} \cdot \mathbf{x}_0$ is

$$y_{-2}' = m_{-2,-1} g_{-1,0}$$
$$y_{-1}' = m_{-1,-1} g_{-1,0} + m_{-1,0} g_{0,0}$$
$$y_0' = m_{0,-1} g_{-1,0} + m_{0,0} g_{0,0} + m_{0,1} g_{1,0} \quad . \tag{16}$$
$$y_1' = m_{1,1} g_{1,0} + m_{1,0} g_{0,0}$$
$$y_2' = m_{2,1} g_{1,0}$$

Using $m_{k(k+1)} = -m_{k(k-1)} = m$, $m_{l(l\pm 1)} = m_{k(k\pm 1)}$ and $m_{kk} = m_{ll}$, we obtain $\Delta \mathbf{y} = (\mathbf{G} \cdot \mathbf{M} - \mathbf{M} \cdot \mathbf{G}) \cdot \mathbf{x}_0$, where

$$\Delta y_{-2} = m(g_{-2,-1} - g_{-1,0})$$
$$\Delta y_{-1} = m(g_{-1,-1} - g_{0,0})$$
$$\Delta y_0 = m(g_{0,-1} + g_{-1,0} - g_{1,0} - g_{0,1}) \quad . \tag{17}$$
$$\Delta y_1 = -m(g_{1,1} - g_{0,0})$$
$$\Delta y_2 = -m(g_{2,1} - g_{1,0})$$

From here, we can attribute the arising of the noncommutativity to two aspects of the optomechanical coupling matrix of our modulator: 1) unlike the broadband EOM, the transmission through the modulated nanophotonic cavity at the center frequency $g_{0,0}$ is significantly different than that from the sidebands ($g_{1,1}$ and $g_{-1,-1}$), i.e. the non-unitarity, associated with the synthetic lattice of our resonating acousto-optic modulator, therefore $\Delta y_{\pm 1}$ are generally nontrivial and are more pronounced at larger transmission difference; 2) as we have explained in Section 8, the two-way frequency conversions between a pair of sidebands ($g_{kl}$ and $g_{lk}$) are highly phase-anisotropic and have a strong dependence on the driving phase $\phi$, which result in non-vanishing contrast at the center-frequency component ($\Delta y_0$) of the output. While our demonstrations involve a broadband and a resonating phase modulators, we note that these two factors ensure the sufficient and necessary conditions for the noncommutativity of the cascaded phase modulator group with the full parametric space of $(\beta, \Omega, \phi, \omega_0, \kappa, \kappa_{ex})$.